\documentclass[fleqn,usenatbib]{mnras}

\usepackage{lipsum}
\usepackage{mwe}

\usepackage{lipsum,graphicx}
\usepackage{newtxtext,newtxmath}
\usepackage[T1]{fontenc}
\usepackage{ae,aecompl}
\usepackage{adjustbox}
\usepackage{amsfonts}
\usepackage{amsmath}
\usepackage{graphicx}	
\usepackage{tabularx}
\usepackage{verbatim}
\usepackage{xspace}
\usepackage{adjustbox}
\usepackage{rotating, lipsum, babel}
\usepackage{MnSymbol}

\usepackage{subfig}

\newcommand{\pc}{\,\mathrm{pc}}
\newcommand{\Msun}{\,\mathrm{M}_{\odot}}
\newcommand{\kpc}{\,\mathrm{kpc}}
\newcommand{\Gyr}{\,\mathrm{Gyr}}

\newcommand{\kms}{\,\mathrm{km\,s}^{-1}}

\newcommand{\rh}{r_\mathrm{h,i}}
\newcommand{\RG}{R_\mathrm{G}}
\newcommand{\dens}{\rho_\mathrm{h,i}}

\newcommand{\Md}{M_\mathrm{d}}

\newcommand{\Nbody}{$N$-body\xspace}

\newcommand{\secref}[1]{Section~\ref{#1}}
\newcommand{\figref}[1]{Figure~\ref{#1}}
\newcommand{\tabref}[1]{Table~\ref{#1}}

\newcommand{\frem}{f_\mathrm{remn}}
\newcommand{\floss}{f_\mathrm{Mloss}}
\newcommand{\fBH}{f_\mathrm{BH}}
\newcommand{\fWD}{f_\mathrm{NS+WD}}

\graphicspath{{Figures/}}

\title[Impact of BH Natal Kicks on Stellar Remnant Fraction in GCs]{Origin of High Dark Remnant Fractions in Milky Way Globular Clusters: The Crucial Role of Initial Black Hole Retention}

\author[A. Rostami-Shirazi et al.]{
Ali Rostami-Shirazi$^{1}$\thanks{E-mail: a.rostami@iasbs.ac.ir},Holger Baumgardt$^{2}$, Akram Hasani Zonoozi$^{1}$, S. Mojtaba Ghasemi$^{1}$, \newauthor and Hosein Haghi $^{1}$
\\
$^{1}$Department of Physics, Institute for Advanced Studies in Basic Sciences (IASBS), 444 Prof. Sobouti Blvd., Zanjan 45137-66731, Iran\\
$^{2}$School of Mathematics and Physics, The University of Queensland, St Lucia, QLD 4072, Australia\\
}

\date{Accepted XXX. Received YYY; in original form ZZZ}

\pubyear{2024}

\begin{document}
\label{firstpage}
\pagerange{\pageref{firstpage}--\pageref{lastpage}}
\maketitle

%%%%%%%%%%%%%%%%%%%%%%%%%%%%%%%%%%%%%%%%%%%%%%%%%%
\begin{abstract}

Comparing the dynamical and stellar masses of Milky Way (MW) globular clusters (GCs) reveals a discrepancy exceeding a factor of two. Since this substantial invisible mass is concentrated in the cluster centre, it is attributed to stellar remnants. The majority of mass in remnants consists of white dwarfs (WDs). Allocating over half of a GC's current mass to WDs could significantly restrict the dynamical evolution scenarios governing stellar clusters. As the most massive stars in GCs, black holes (BHs) exert a substantial effect on the escape rate of lower mass stars, such as WDs. This paper aims to identify which scenarios of BH natal kicks can accurately reproduce the notable dark remnant fraction observed in MW GCs. We compare the observed remnant fraction of MW GCs with a comprehensive grid of direct \Nbody simulations while adjusting the natal kick received by BHs. Our results reveal that simulations employing low natal kicks to BHs are the only ones capable of mirroring the remnant fraction of MW GCs. According to the Spitzer instability, the presence of a BH population prompts the formation of a BH sub-system (BHSub) at the centre of a star cluster. The BHSub serves as an energetic power plant, continually releasing kinetic energy through few-body encounters between single and binary BHs, and transferring the generated energy to the entire stellar population. This energy induces a significant difference in the ejection rate of stellar remnants and luminous stars, ultimately increasing the fraction of dark remnants within the star cluster.

\end{abstract}


%%%%%%%%%%%%%%%%%%%%%%%%%%%%%%%%%%%%%%%%%%%%%%%%%%%%%%%%%%%%%%%%%%%%%%%%%%%%%
\section{Introduction}\label{sec:intro}

An enduring challenge in astrophysics, spanning nearly a century, revolves around the persistent disparity between mass estimation inferred from the kinematics of stars within stellar systems and the associated luminosities \citep{Zwicky1933,Rubin1970,Ostriker1973,Faber1979,Persic1992,McGaugh2000,McGaugh2005,Walker2009,McGaugh2014}. The baryonic mass within stellar systems, derived from the conversion of luminosity into mass, consistently reveals substantially smaller values than assessments derived from kinematic considerations (see \citealt{Salucci2019,Arbey2021} for recent reviews). In this context, globular clusters (GCs) were thought to stand out as a noteworthy anomaly, primarily justified by their consistently low dynamical mass-to-light ratios \citep{McLaughlin2005,Strader2009,Strader2011}.

\citet{Gunn1979} found evidence for dark mass in the GC M3 and concluded that it is compatible with the amount of remnants produced by stellar evolution. \citet{Heggie1996} revealed that about half of the mass in NGC 6397 and 47 Tuc is unobservable. \citet{Gebhardt1997} suggest that invisible mass constitutes the majority of the mass in GC M15. \cite{Sollima2012} identified signs of considerable invisible mass within GCs. They determined the dynamical and luminous masses of a sample of six Galactic GCs, taking into account the effects of mass segregation, anisotropy, and unresolved binaries. Their approach involved simultaneous fitting of luminosity functions and line-of-sight velocity dispersion profiles of GCs with multimass analytical models, treating the present-day mass function as a free parameter. The analysis revealed a remarkable discrepancy between stellar and dynamical masses, with stellar masses being lower than the dynamical masses by an average of $\sim 40$ per\,cent.

\cite{Sollima2016} presented a notable improvement in the observational estimation of the fraction and distribution of dark mass for GCs examined in their earlier study, \cite{Sollima2012}. Their analysis successfully revealed the radial distribution of dark mass for NGC 6218 and NGC 288. They found that the invisible mass constitutes more than 60 per\,cent of the total mass within the radius less than 1.6 arcmin in both GCs. They excluded the possibility that the elevated dark mass fractions result from spurious effects caused by binaries and tidal heating. Examining the radial distribution of dark mass, they inferred that the discrepancy between luminous and dynamical mass is particularly heightened in the central regions of the clusters. Indeed, the dark component appears to be segregated in the cluster core.

In the context of the nature of this dark mass, which typically constitutes over half of the GC total mass, \cite{Sollima2016} put forth three hypotheses: 1) the presence of an intermediate-mass black hole (IMBH) in the centre of the GCs, 2) a modest amount of non-baryonic dark matter, and 3) an unexpectedly high fraction of retained stellar remnants. They showed that the mass of the central IMBH depends on the fraction and distribution of other remnants. Indeed, the larger the fraction of remnants, the smaller the mass of the hypothetical IMBH. Moreover, they assessed the probability of the dark component within the cluster being associated with dark matter as very low. This evaluation is grounded in \Nbody simulations that assumed GCs were surrounded by cored dark matter haloes. These simulations predict that due to interactions with cluster stars and tidal stripping by the host galaxy's field, the central parts of GCs might be left relatively poor in dark matter, while it is either distributed at the outermost regions of the cluster or tidally cut off \citep{Mashchenko2005,Penarrubia2008,Baumgardt2008UCD}. This contradicts the observed high concentration of dark mass in GCs. Finally, \cite{Sollima2016} concluded that the most likely interpretation associates the substantial invisible mass within the GCs with a higher-than-anticipated fraction of retained stellar remnants.

Over the past few years, many observational analyses aimed at determining the mass fraction of dark stellar remnants ($\frem$) in GCs. Employing a methodology described in \cite{Sollima2012}, \cite{Sollima2017} determined $\frem$ for a subset of 29 Milky Way (MW) GCs. They found the dynamical mass of the GCs to be, on average, nearly twice their luminous mass ($\frem\sim0.5$). Furthermore, a significant correlation was detected between the present-day mass function slopes and $\frem$ within the GCs. Recently, \citet{Dickson2023} acquired total and visible mass components for 34 MW GCs. In agreement with \cite{Sollima2017}, they determined that, on average, the dark stellar remnants constitute around 54 per\,cent of the total mass of the GCs. They also identified a correlation between $\frem$ and the dynamical age of GCs. Moreover, \cite{Ebrahimi2020} examined a sample of 31 MW GCs, determining $\frem$ through a combination of available radial velocity information and deep photometric data. They identified 28 GCs with a remnant fraction exceeding 40 per\,cent, including 23 with a remnant fraction surpassing 50 per\,cent.

The final phase of the stellar evolution process results in the formation of stellar remnants, including white dwarfs (WDs), neutron stars (NSs), and black holes (BHs), whose luminosities are several orders of magnitude smaller than those of main-sequence stars. The $\frem$ in clusters is governed by the combined effects of stellar and dynamical evolution in these stellar systems. As the most massive stars in GCs, BHs play a crucial role in the dynamical evolution of GCs \citep{Breen2013,Longwang2020,Gieles2023,Rostami2024,Ghasemi2024}. The extent of natal velocity kicks that BHs receive during their formation in a supernova explosion is still a matter of debate, leading to uncertainties about the initial number of BHs remaining in star clusters. It was initially believed that BHs experience a significant natal kick, which could accelerate them beyond the escape velocity, resulting in the exodus of almost all BHs from the cluster as soon as their formation. However, over the last decades, there has been a significant change in our understanding of BH retention in GCs. Several observational studies have provided evidence of the presence of individual BHs in many GCs \citep{Maccarone2007,Shih2010,Barnard2011,Strader2012,Chomiuk2013,Miller-Jones2015,Giesers2018,Giesers2019,Saracino2022,Saracino2023}.

On the theoretical and computational side, several studies have shown that retaining BHs within certain clusters is necessary for reproducing their observable properties \citep{banerjee2011,Morscher2015,Peuten2016,Baumgardt2019,Zocchi2019,Kremer2019,Weatherford2020,Gieles2021,Torniamenti2023}. In this context, \citet{Rostami2024-II} revealed that a high initial retention of BHs (low natal kicks) can explain the difference of over 50 per\,cent in half-mass and half-light radii between metal-poor and metal-rich MW GCs. Additionally, the absence of significant natal kicks of BHs can account for the lack of metal-poor GCs in the inner regions of the MW \citep{Rostami2024-II}. This is because the heavier and less expanded BH sub-system (BHSub) in metal-poor clusters leads to more intense few-body encounters, thereby injecting greater kinetic energy into the entire stellar population. Consequently, they experience larger expansion and higher evaporation rates than metal-rich clusters. The higher energy production within the BHSub of lower-metallicity GCs causes them to dissolve prior to a Hubble time in proximity to the Galactic center.

In order to estimate the present-day BH mass fraction within MW GCs, recent studies have employed best-fitting multimass models, evaluations of visible mass segregation, and assessments of the central surface brightness of several GCs, indicating that a large number of the GCs are consistent with hosting little or no BHs \citep{Askar2018,Weatherford2020,Dickson2023}. Unlocking the mystery of the magnitude of BH natal kicks might lie in the inferred present-day fraction of BH masses. In \citet{Rostami2024}, we conducted a series of direct \Nbody simulations and concluded that, even with the assumption of zero natal kicks for BHs, the majority of MW GCs (87 per cent), experience the near complete depletion of the BH population through few-body encounters (assuming MW GCs followed canonical initial mass functions). For most of the remaining GCs, around 85 per\,cent of BHs are ejected from the cluster up to the present day. Our results highlighted that achieving the present-day BH mass fraction doesn't require BHs to receive a high natal kick. Instead, even with high initial retention of BHs, a substantial number of them are depleted through few-body encounters, shaping the present-day BH mass fraction of MW GCs.

In this paper, we aim to explore the impact of the natal kick received by BHs on the $\frem$ of star clusters. The main question that we ultimately aim to answer is which of the evolutionary scenarios for star clusters can accurately reproduce the notable $\frem$ observed in MW GCs: the presence of a segregated sub-system of BHs within GCs due to low natal kicks of BHs or, alternatively, the removal of BHs immediately after their formation from GCs through a significant natal kick. To this end, we compare the $\frem$ for a subsample of 29 MW GCs determined by \cite{Ebrahimi2020} with a comprehensive grid of direct \Nbody simulations of star clusters over a wide range of initial half-mass radii ($\rh$) and Galactocentric distances ($\RG$), while varying the natal kick of BHs. Our paper is organized as follows: In \secref{sec:methodology}, we describe the initial setup of the \Nbody models and our simulation method. The main results are presented in \secref{sec:results}, where we compare the $\frem$ of simulated clusters in the presence (\secref{sec:Kick}) and absence (\secref{sec:NKick}) of the BHs' natal kick with the observed $\frem$ in Galactic GCs. Finally, \secref{sec:conclusion} consists of a discussion and the conclusions.

%%%%%%%%%%%%%%%%%%%%%%%%%%%%%%%%%%%%%%%%%%%%%%%%%%%%%%%%%%%%%%%%%%%%%%%%%%%%%

\section{DESCRIPTION OF THE MODELS AND INITIAL CONDITIONS}\label{sec:methodology}

\begin{table}
    \caption{Initial parameters of simulated clusters classified based on the kick velocities assigned to BHs in each model. Check marks indicate the natal kick scenarios employed for the model characterized by $\rh$ (column 2) and $\RG$ (column 3).}
	\centering
	\begin{tabular}{ccccccc}
		
  		\hline 
		
            \hline 
		Model & $ \rh $ & $ \RG $ & $\sigma_\mathrm{BH}=190$ & $\sigma_\mathrm{BH}=0$ \\ 
		& $(\pc)$ & $(\kpc)$ & $(\kms)$ & $(\kms)$ \\
 
		\hline
  
        M1  & 0.8 & 8 &  & $\checkmark$\\
        M2  & 1 & 2 & $\checkmark$ &$\checkmark$ &\\
        M3  & 1 & 3 & $\checkmark$ &$\checkmark$ &\\
        M4  & 1 & 4 & $\checkmark$ & $\checkmark$\\
        M5  & 1 & 8 & $\checkmark$ & $\checkmark$\\
        M6  & 1 & 12 & $\checkmark$ & $\checkmark$\\
        M7  & 3 & 2 & $\checkmark$ & $\checkmark$\\
        M8  & 3 & 3 & $\checkmark$ & $\checkmark$\\
        M9  & 3 & 4 & $\checkmark$ & $\checkmark$\\
        M10  & 3 & 6 & $\checkmark$ & $\checkmark$\\
        M11  & 3 & 8 & $\checkmark$ & $\checkmark$\\
        M12  & 3 & 12 & $\checkmark$ & $\checkmark$\\
        M13  & 3 & 16 & $\checkmark$ & $\checkmark$\\
        M14  & 5 & 2 & $\checkmark$ & $\checkmark$\\
        M15  & 5 & 3 & $\checkmark$ & $\checkmark$\\
        M16  & 5 & 4 &  & $\checkmark$\\
        M17  & 5 & 6 &  & $\checkmark$\\
        M18  & 5 & 8 & $\checkmark$ & $\checkmark$\\
        M19  & 5 & 12 & $\checkmark$ & $\checkmark$\\
        M20  & 5 & 16 & $\checkmark$ & $\checkmark$\\
        
	\hline      

        \hline
        
	\end{tabular}
	\label{tab:initial_conditions}
\end{table}

\par The natal kicks received by BHs are, to date, poorly constrained and understood from both observational and theoretical points of view. The post-formation mass fraction of BHs in the cluster is dictated by the magnitude of their natal kicks. In scenarios where the natal kick is high, BHs are ejected from the cluster immediately after formation, whereas low natal kicks lead to the gradual depletion of BHs through few-body encounters. Comparing the rapid ejection of BHs from the cluster due to natal kicks with their gradual depletion caused by few-body encounters may result in significant distinctions in the dynamic evolution of clusters and the fraction of dark remnants retained in them.

To quantify the effect of the natal kick received by BHs on the $\frem$ of star clusters, we utilize the collisional \Nbody code NBODY7 \citep{Aarseth2012} which is an enhanced version of the widely used NBODY6 direct \Nbody evolution code \citep{Aarseth2003, Nitadori2012}. The state-of-the-art NBODY6 code provides an extensive treatment of both single and binary stellar evolution from the zero-age main sequence to their final stages, integrating the SSE/BSE algorithms and analytical fitting functions established by \citet{hurley2000}. NBODY7 incorporates significant updates in two key facets of the single stellar evolution process. First, it determines the masses of compact objects using the methodology proposed by \citet{Belczynski2008}. Second, the code implements the model presented by \citet{vink2001} to account for mass loss due to stellar winds, as detailed in \citet{Belczynski2010}. NBODY7 employs the Algorithmic Regularization Chain introduced by \citet{mikkola1999} to implement high-precision and efficient methods for managing strong gravitational encounters, including binary stars, instead of the classic Chain Regularization approach used in NBODY6 \citep{Mikkola1993}. Utilizing this enhanced algorithmic regularization method for compact subsystems enables the inclusion of general relativistic effects via post-Newtonian terms and realistic parameters. Moreover, this method provides a more thorough and reliable treatment of dynamically forming multiple systems in dense environments,  especially those containing massive objects such as BHs. This family of codes is compatible with graphic processing units (GPUs), which we leveraged to substantially reduce the simulation runtime. Our simulations were conducted on a desktop workstation equipped with Nvidia 1080 GPUs at the Institute for Advanced Studies in Basic Sciences (IASBS).

We conducted simulations with an initial cluster mass of  $3\times10^4 \Msun$ and let the models evolve for $13 \Gyr$. The clusters are populated with stars in the mass range of $m_{\mathrm{low}}=0.07\Msun$ and $m_{\mathrm{up}}=150\Msun$, following the Canonical initial mass function \citep{Kroupa2001}. Stars are spatially distributed according to the Plummer density profile \citep{Plummer1911}, and initially, the cluster is in virial equilibrium. Our modelled clusters are not initially mass-segregated. Additionally, primordial binaries are not considered for simplicity and to reduce computational costs, although binaries are regularly created and destroyed via three-body interactions. The adopted stellar metallicity is $0.005$ ($\sim$ 0.25 $Z_{\sun}$). The initial orbital velocities of the clusters are set in such a way to keep them in a circular orbit through the host Galaxy at their distances of $\RG$, which range from $2$ to $16\kpc$ for different clusters. The clusters are exposed to a static Galactic potential consisting of three components: a point-mass bulge, a Miyamoto-Nagai disc \citep{Miyamoto1975}, and a logarithmic halo \citep{miholics2014}, scaled such that the circular velocity at $8.5\kpc$ is $220\kms$. The bulge is modelled as a central point mass with a mass of $1.5\times10^{10}\Msun$, while the disc is characterized by a scale length of $4 \kpc$ and a height of $0.5 \kpc$ \citep{Read2006}, with a corresponding mass of $\Md = 5\times10^{10}\Msun$ as suggested by \citet{Xue2008}.

The natal kick received by BHs is adopted as a free parameter in our simulations. We considered two scenarios for the BH natal kick in our simulated models: 1) Following their formation, all BHs receive a velocity kick drawn from a Maxwellian distribution with a one-dimensional dispersion of $\sigma_{\mathrm{kick}} = 190 \kms$ \citep{Hansen1997}. This results in the expulsion of nearly all BHs from the cluster upon formation. 2) BHs are devoid of any natal kick, thus ensuring their initial retention within the cluster. It should be noted that assuming a 100 per\,cent initial retention fraction for BHs may seem unrealistic. A common model \citep{Belczynski2008, Fryer2012,Banerjee2020} for supernova natal kick magnitude assumes NS-like kicks \citep{Hobbs2005} for BHs as well, but which are scaled down linearly with an increasing material fallback fraction, the so-called  canonical supernova kicks, upon which mass fallback typically results in about half of BHs being ejected at birth \citep{Chatterjee2017}. However, in this paper, we adopt this extreme limit for the retained BHs in modeled clusters, to isolate the pure effect of the initial BH mass fraction on the dark remnant fraction of clusters.

The formation of WDs marks the final evolutionary phase for lower-mass stars. Due to the extended lifetimes of WD progenitors, the fraction of WDs undergoes a gradual rise over the cluster's lifetime, positioning WDs as a significant contributor to the dark mass fraction of GCs. During post main-sequence evolution, non-spherically symmetric mass loss leads to an isotropic recoil speed in the WD remnant \citep{Fellhauer2003}. The magnitude of WDs natal kick can affect the fraction of dark remnant within the clusters. Recently, \cite{El-Badry2018} found breaks in the separation distribution of main-sequence star-WD and WD–WD binaries, using $Gaia$ DR2 data. They showed that these breaks could be explained if WDs incur a natal kick with a typical magnitude of $0.75\kms$ ($\sigma_{\mathrm{WD}}=0.5\kms$). Their findings rule out typical kick velocities above $2\kms$ \citep{Heyl2007,Heyl2008a,Heyl2008b}. Thus, in our simulations, we applied kick velocities with a dispersion of $\sigma_{\mathrm{WD}} = 0.5\kms$ for WDs.

In total, for our modelled clusters, we carried out two sets of simulations based on varying natal kicks applied to BHs. An overview of the initial parameters of modelled clusters and performed sets of simulations are given in \tabref{tab:initial_conditions}.

%%%%%%%%%%%%%%%%%%%%%%%%%%%%%%%%%%%%%%%%%%%%%%%%%%%%%%%%%%%%%%%%%%%%%%%%%%%%%
\section{Results}\label{sec:results}

In order to study the influence of BH retention on the $\frem$ in star clusters, we prepared two sets of models, based on the presence and absence of BH natal kicks. Both sets will be discussed in the following sections. In the models presented in \secref{sec:Kick}, BHs receive a kick velocity drawn from a Maxwellian distribution with a one-dimensional dispersion of $\sigma_\mathrm{BH} = 190 \kms$. Subsequently, \secref{sec:NKick} contains models where BHs experience no natal kick, $\sigma_\mathrm{BH} = 0 \kms$. In each section, we compare the $\frem$ of modelled clusters with that of MW GCs. Our primary goal is to identify the optimal scenarios for natal kick that accurately reproduce the observed remnant fraction of MW GCs.

Since WDs constitute the predominant portion of the dark stellar remnants in star clusters, a successful scenario involves preserving a significant fraction of WDs throughout cluster evolution. Assuming stars within the mass range of 0.85-7.5 $\Msun$ as the progenitors of WDs \citep{Cummings2018}, about $30$ per\,cent of the initial mass in our modelled clusters consists of stars capable of evolving into WDs. Indeed, a first-order estimation indicates that approximately 30 per\,cent of the cluster's initial mass can be allocated to WDs. Our estimation reveals a substantial contrast between the initial mass fraction of WDs and their current dominance, which exceeds 50 per\,cent of the cluster's mass. To drive the fraction of WDs from 30 per\,cent to over 50 per\,cent during cluster evolution, the evaporation rate of luminous stars, i.e., the nuclear-burning stars, must significantly surpass the escape rate of WDs within the cluster. This process is expected to result in a correlation between the WD mass fraction and the mass function slope. Therefore, a scenario that reproduces the observed $\frem$ in MW GCs must introduce a notable discrepancy in the expulsion rates between WDs and luminous stars within the cluster.

%Note that this estimate does not account for the mass loss experienced by WD progenitors. When including this mass loss, a smaller fraction of the cluster's total mass can be attributed to WDs.

%_%_%_%_%_%_%_%_%_%_%_%_%_%_%_%_%_%_%_%_%_%_%_%_%_%_%_%_%_%_%_%_%_
\subsection{Presence of BH natal velocity kicks}\label{sec:Kick}

\begin{figure}
  \centering
  \includegraphics[width=\linewidth]{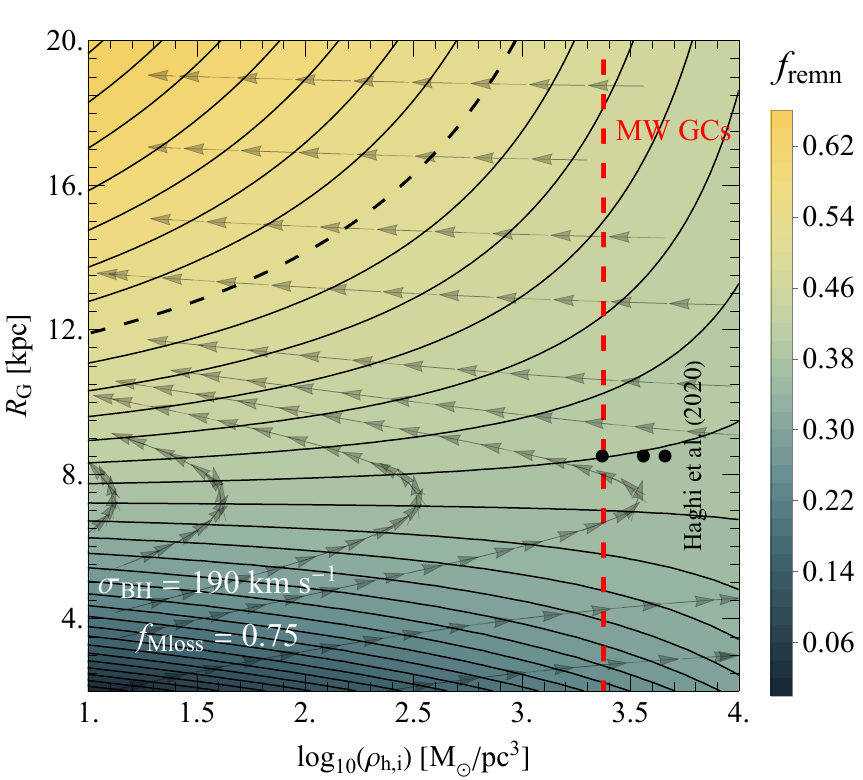}
  \caption{Dark remnant mass fraction of the modelled clusters with $\sigma_\mathrm{BH}=190\kms$ at the moment when they have lost 75 per cent of their mass since formation. $\frem$ is shown as a colour-coded plot within the phase space of $\log _{10}(\dens)$ and $\RG$. The colour bar represents higher fractions by warmer colours and lower fractions by cooler colours. The right side of the perpendicular with a red dashed line shows an acceptable range for the initial density of the MW GCs. The dashed contour represents $\frem=0.5$. Dim black field vectors are superimposed on the plot, indicating the gradient of $\frem$ with respect to $\log _{10}(\dens)$ and $\RG$. The direction of these vectors shows how $\frem$ changes with varying $\dens$ and $\RG$. Rightward (leftward) vectors indicate increasing (decreasing) $\frem$ with $\dens$, and upward components show increasing $\frem$ with $\RG$.}
  \label{fig:BH190-DW05-23-75per}
\end{figure}

Applying a high natal kick ($\sigma_{\mathrm{kick}} = 190 \kms$) to massive stellar remnants, almost all BHs and NSs are ejected from the cluster within the first 100 Myr. Following the ejection of BHs and NSs, WDs are the only dark remnants remaining in the cluster, with their mass fraction mirroring the cluster's dark remnant fraction. Actually, the interplay between the escape rate of WDs and the evaporation rate of luminous stars determines the $\frem$ of the cluster. A slower WD expulsion rate than the evaporation rate of luminous stars results in a higher $\frem$ in the cluster.

The initial density and orbital parameters of a star cluster exert considerable influence on the evaporation rate of luminous stars and the escape rate of stellar remnants \citep{Rostami2024}. Consequently, the $\frem$ of the cluster is expected to depend on both the initial density and orbital radius of the cluster. To examine this correlation, we conducted a series of simulations using modelled clusters characterized by varying initial density and $\RG$ (\tabref{tab:initial_conditions}). Given that all the modelled clusters have the same initial mass of $3\times10^4 \Msun$, adjusting the value of $\rh$ leads to varying initial densities within the half-mass radius, $\dens = 3 M_{\mathrm{cluster}}/{8 \pi  r_\mathrm{h,i}^3}$. We calculate models with initial half-mass radii of $\rh(\pc)\in\{1, 3, 5\}$, moving on circular orbits at Galactocentric distances ranging from $\RG=2$ to 16 $\kpc$.

\citet{Dickson2023} identified a correlation between $\frem$ and the dynamical age of GCs. Hence, we analyze the $\frem$ of modelled clusters and MW GCs at an equivalent fraction of mass lost by the cluster ($\floss$). The MW GCs have, on average, lost approximately 75 per\,cent of their initial stellar masses ($\floss=0.75$) due to stellar evolution or prolonged dynamical processes \citep{Webb2015,Baumgardt2017-75per,Baumgardt2019mean}. To initiate our analysis, we determine the $\frem$ for our modelled clusters when 75 per\,cent of their initial mass has been lost. Subsequently, we derived the best-fitting function for $\frem$ in dependence of $\log _{10}(\dens)$ and $\log _{10}(\RG)$ of the modelled clusters in the following mathematical form (following our approach in \citealt{Rostami2024}):
\begin{equation}\label{eq:general}
    \frem\left(\RG,\dens \right)=a (\dens)  \log _{10}\left(\frac{\RG}{\text{kpc}}\right)+b (\dens),  
\end{equation}
where $a$ and $b$ represent the best-fitting parameters which are themselves a function of $\log _{10}(\dens)$, defined as:
\begin{equation}\label{eq:a-b}
    \left\{a,b\right\} (\dens) =\left\{a_{1},b_{1}\right\}\log _{10}\left(\frac{\dens}{\mathrm{M}_{\odot }\text{pc}^{-3}} \right)+\left\{a_{2},b_{2}\right\}. 
\end{equation}

\figref{fig:BH190-DW05-23-75per} illustrates $\frem$ as a function of $\log _{10}(\dens)$ and $\RG$ for clusters with a kick velocity set to $\sigma_{\mathrm{BH}} = 190 \kms$. This plot reveals a trend where $\frem$ increases with increasing $\RG$. It is well known that as the $\RG$ of an orbiting cluster increases, which means reducing the tidal field of the host galaxy, the mass-loss rate of the cluster decreases \citep{Baumgardt2003}. The reduced mass-loss rate in the cluster provides sufficient time for massive stars to segregate towards the cluster centre and become immersed in its central potential. Concentrating massive stars in the cluster's centre reduces the escape rate of WDs and ensures that high-mass luminous stars undergo their final evolutionary stages inside the cluster. Indeed, they transition out of their main-sequence phases and evolve into remnants, thereby increasing the population of WDs. Therefore, the mass fraction of WDs within the cluster increases as $\RG$ increases.

\figref{fig:BH190-DW05-23-75per} displays that within Galactic radii $\RG<8\kpc$, where the Galactic tidal force plays a more substantial role in the dynamical evolution of star clusters, an increase in the initial density leads to a rise in $\frem$. Since WDs emerge as the predominant high-mass components in the cluster, dynamical friction leads to their gradual segregation towards the cluster centre. As the $\dens$ of the cluster increases, there is a consequential decrease in the half-mass relaxation time. Consequently, the segregation timescale for high-mass cluster members, including WDs, decreases. When WDs segregate towards the central part of the cluster, the luminous stars in the cluster halo shield them from the Galactic tidal force. Thus, an accelerated segregation of WDs leads to a reduction in their escape rate from the cluster. Indeed, within $\RG<8\kpc$, the escape rate of WDs decreases more steeply than the evaporation rate of luminous stars as density increases, thereby enhancing $\frem$ in the cluster.

This trend converses for clusters situated at greater orbital distances. In high-density models, the compact accumulation of WDs sparks intense dynamical activity and catalyzes encounters between WDs through few-body interactions. The high frequency of these encounters accelerates the self-depletion of WDs. This outcome signifies a lower fraction of WDs within models characterized by high initial densities. The process of segregating and concentrating WDs to enable high-energy collisions is time-consuming. Consequently, it is only in weak tidal fields that clusters stand a chance of activating their WD populations dynamically, leading to an accelerated rate of WD self-depletion. Hence, in outer Galactic regions ($\RG>8\kpc$), the mass fraction of WDs declines as density increases.

\figref{fig:BH190-DW05-23-75per} effectively illustrates that the interplay among three key mechanisms—tidal evaporation, mass segregation, and self-depletion of WDs through few-body interactions—governs the contour of $\frem$ in the $\log _{10}(\dens)$-$\RG$ parameter space. To further elucidate these trends, we have superimposed dim black field vectors on the \figref{fig:BH190-DW05-23-75per}, representing the gradient of $\frem$ with respect to $\log _{10}(\dens)$ and $\RG$. The orientation of these vectors provides valuable insights into the dominant mechanisms across different regions of the parameter space. For $\RG<8\kpc$, the vectors incline to the right, indicating a positive correlation between $\frem$ and $\dens$. This trend corroborates our earlier discussion on enhanced mass segregation and retention of WDs in denser clusters within strong tidal fields. Conversely, for $\RG>8\kpc$, the vectors tilt to the left, signifying a negative correlation between $\frem$ and $\dens$. This leftward orientation corroborates our explanation of accelerated WD self-depletion in high-density clusters situated in weak tidal fields. Notably, all field vectors exhibit a slight vertical component, consistently pointing upward. This vertical orientation confirms that $\frem$ generally increases with increasing $\RG$ across all density ranges, reinforcing our interpretation of reduced tidal stripping and enhanced WD retention at larger Galactocentric distances.

\begin{figure}
  \centering
  \includegraphics[width=0.9\linewidth]{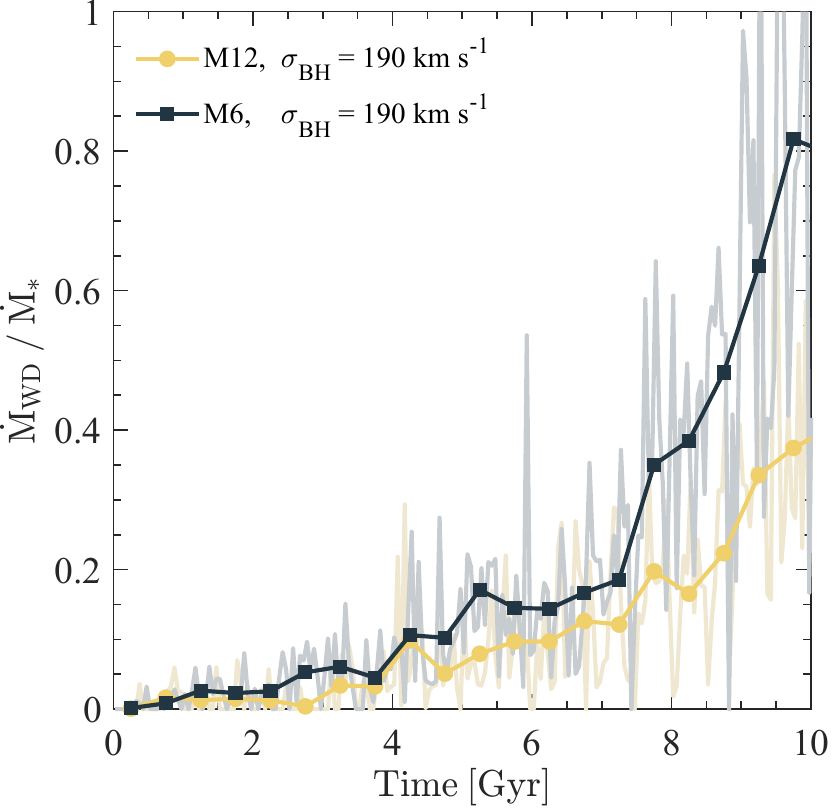}
  \caption{The time evolution of the ratio of WDs' escape rate to the outflow mass rate of luminous stars within clusters, all featuring identical kick velocities ($\sigma_\mathrm{BH}=190\kms$), but varying in $\rh$. Yellow circles denote the M12 model ($\rh=3 \pc$), while black squares indicate the M6 model ($\rh=1 \pc$). Bolder-coloured lines indicate the mean values of $\overset{.}{M}\mathrm{WD} / \overset{.}{M}{*}$ for the fainter (same-coloured) lines in the background.}
  \label{fig:dMWD_out_dMstar_time_rh06_rh2}
\end{figure}

\figref{fig:dMWD_out_dMstar_time_rh06_rh2} illustrates the temporal evolution of the escape rate of WDs divided by the luminous stars' evaporation rate ($\overset{.}{M}_\mathrm{WD} / \overset{.}{M}_{*}$) for modelled clusters at $\RG = 12\kpc$ with $\rh(\pc)\in\{1, 3\}$, referred to as models M6 and M12, respectively. The M6 model exhibits a higher $\overset{.}{M}_\mathrm{WD} / \overset{.}{M}_{*}$ ratio due to heightened encounters among its WDs compared to the M12 model. This diminishes the mass fraction of WDs in the M6 model. Following $10\Gyr$, this ratio in M6 reaches twice that of M12.

Based on the \citet{Marks2012} relation:
\begin{equation}\label{eq:marks}
    \rh=0.1 \times \left(\frac{M_{\text{cluster}}}{\Msun}\right){}^{0.13} \ \text{pc}  
\end{equation}
a cluster with an initial mass of $3\times10^4 \Msun$ possesses an $\rh$ of 0.38$\pc$. It's essential to highlight that our simulations do not consider the gas expulsion effect, thus we must interpret our model's starting conditions as post-gas expulsion. \citet{Baumgardt2007} determined that with a star formation efficiency of approximately $0.3$, massive clusters expand by roughly a factor of $\simeq 3$ during the gas expulsion phase. This expansion factor remains nearly unaffected by the rate of gas removal. Utilizing this expansion factor, the post-gas expulsion density of a cluster with a mass of $3\times10^4 \Msun$ will be $\log _{10}(\dens)=3.4$. We employ this value as a conservative estimate for the initial density of the MW GCs, as indicated by the vertical red dashed line in \figref{fig:BH190-DW05-23-75per}. It's noteworthy that the initial density of the MW GCs typically surpasses what our model indicates. This is because the MW GCs are generally more massive than the modelled clusters we've utilized in our simulations. \figref{fig:BH190-DW05-23-75per} shows that the $\frem$, for modelled clusters with an acceptable range of initial density for the MW GCs, varies from 24 per\,cent in the innermost Galactic regions to about 46 per\,cent at $\RG=20 \kpc$.

Note that the mass fraction of WDs in a cluster depends on the interplay between stellar evolution and dynamical processes. Since the clusters in the lower left corner of \figref{fig:BH190-DW05-23-75per} dissolve quickly and have a notable age difference at $\floss=0.75$ compared to clusters with higher $\RG$ or $\dens$, the significant difference in WD mass fraction is predominantly due to stellar evolution, with only faint effects from dynamical process (as clusters age, the WD fraction increases). As we move away from the lower left corner of \figref{fig:BH190-DW05-23-75per}, the influence of dynamical evolution on the observed trends in WD mass fraction becomes more dominant. For instance, the trend of a decreasing WD mass fraction with increasing $\dens$ at $\RG>8\kpc$, identified as WD self-depletion, is primarily driven by dynamical processes. It should also be mentioned that the difference in WD mass fraction at fixed $\floss$ between simulations with $\sigma_\mathrm{BH}=190\kms$ and $0\kms$ mainly arises from the distinct dynamical evolution pathways of their clusters in the presence and absence of BH natal kick. Indeed, the discrepancy in WD mass fraction between models with $\sigma_\mathrm{BH}=190$ and $0\kms$ at fixed $\floss$ appears to be largely independent of stellar evolution effects, due to the unremarkable age distinctions observed in their clusters at the same $\RG$ and $\dens$.To further disentangle the contributions of stellar evolution and dynamical processes, we present supplementary simulations in the \secref{sec:Appendix} where stellar evolution is disabled. These models help elucidate whether the observed trends in the remnant fraction of WDs in clusters are primarily driven by dynamical processes or stellar evolution.

\begin{figure}
  \centering
  \includegraphics[width=0.9\linewidth]{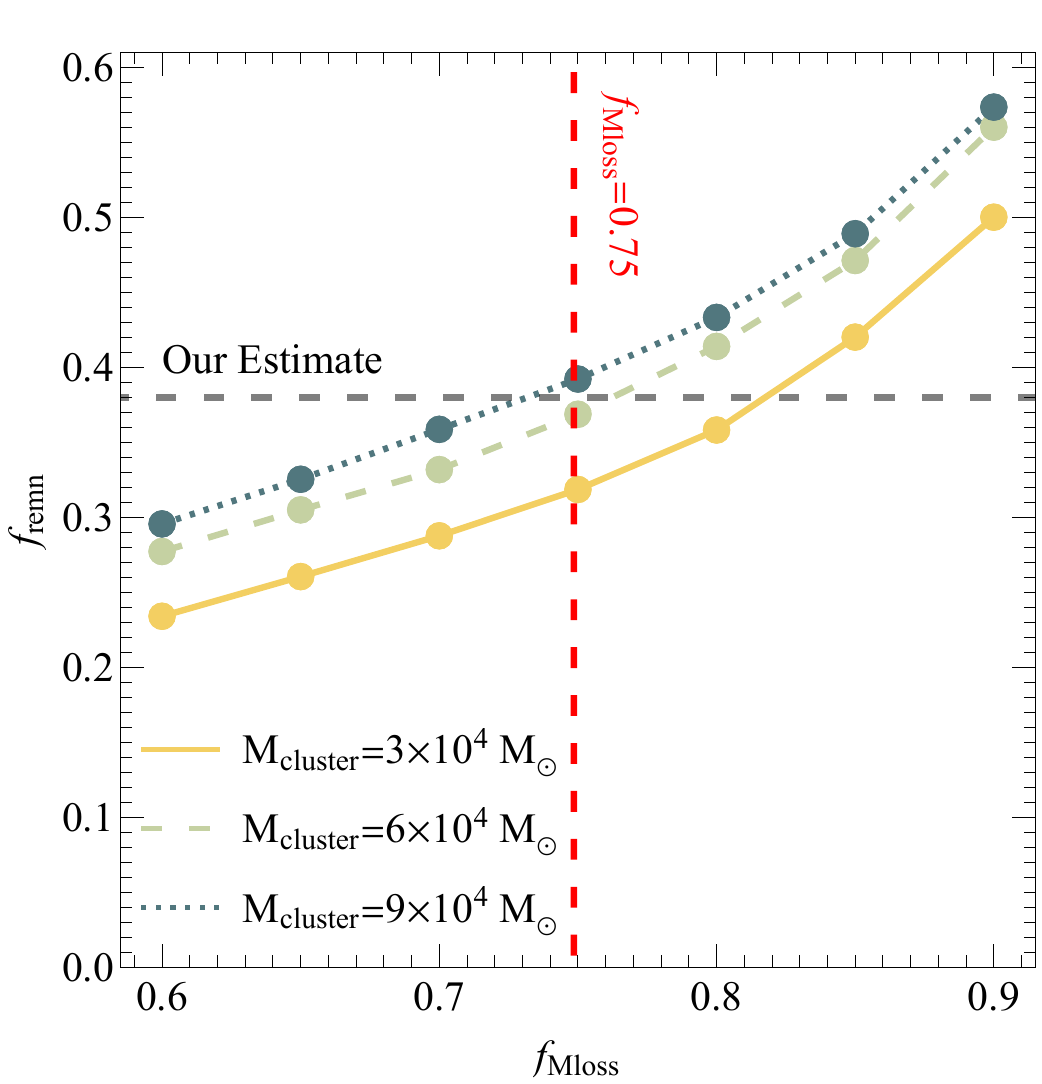}
  \caption{Evolution of $\frem$ across $\floss$ for three selected simulated models with initial masses of $3\times10^4$, $6\times10^4$, and $9\times10^4 \Msun$ from \citet{Haghi2020}. The dashed gray line represents our extrapolation from \figref{fig:BH190-DW05-23-75per}.}
  \label{fig:safaee}
\end{figure}

Our simulations employ an initial cluster mass of $3\times10^4 \Msun$, which is relatively low compared to many MW GCs. In more massive clusters, we anticipate a higher WD mass fraction, primarily due to the extended time required to reach $\floss = 0.75$. This prolonged timeline allows a larger proportion of stars to complete their main-sequence evolution and transition into WDs before the cluster loses a significant portion of its mass. We expect that increases in cluster mass primarily impact $\frem$ through stellar evolutionary effects, while the dynamical mechanisms governing $\frem$ remain relatively invariant across the mass spectrum. We utilized the simulations of \citet{Haghi2020} to assess how initial mass influences $\frem$. Their models encompass a range of initial masses, with $\rh$ following the \citet{Marks2012} relation. We selected models with masses of $3\times10^4$, $6\times10^4$, and $9\times10^4\Msun$, all orbiting at $\RG=8.5\kpc$ in circular orbits. These models, initially gas-embedded, undergo an approximately threefold expansion post-gas expulsion. These models are marked as black dots in \figref{fig:BH190-DW05-23-75per}, noting that all fall within the acceptable initial density range for MW GCs. \figref{fig:safaee} showcases the $\frem$ evolution across $\floss$ values from 0.6 to 0.9. At $\floss=0.75$, the dark remnant fraction increases from 0.32 for the $3\times10^4\Msun$ model to 0.37 and 0.39 for the $6\times10^4$ and $9\times10^4\Msun$ models, respectively. This non-linear progression suggests that the effect of stellar evolution on $\frem$ appears to approach an upper limit, as evidenced by the diminishing returns observed when doubling or tripling the initial mass. Consequently, extrapolating our results to the higher masses typical of MW GCs likely provides a reasonable $\frem$ estimation.

\begin{figure}
  \centering
  \includegraphics[width=\linewidth]{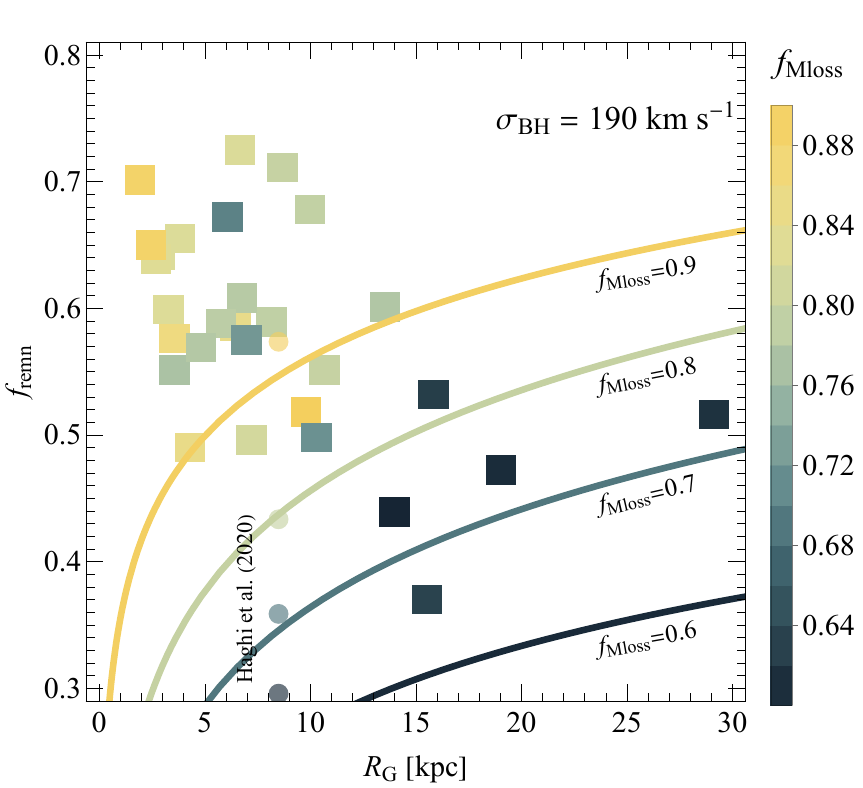}
  \caption{Remnant fraction, $\frem$, of clusters along their Galactic orbits at specific $\floss$ values, obtained for both simulated clusters (represented by curved lines) and MW GCs (depicted by cubes). The colour coding represents the varying degrees of $\floss$, which range from 0.6 to 0.9. The $\frem$ of modelled clusters is depicted for specific mass-loss fractions: $\floss=0.6$, 0.7, 0.8, and 0.9. All simulations were conducted with BH kick velocities of $\sigma_\mathrm{BH} = 190\kms$. Color-coded cubes indicate the corresponding $\floss$ of 29 MW GCs, each paired with the mean $\RG$ and $\frem$ of the respective GC. The color-coded points represent the $\frem$ values of simulated clusters with an initial mass of $9\times10^4 \Msun$ from \citet{Haghi2020}.}
  \label{fig:BH190-DW05-23-tot}
\end{figure}

\citet{Webb2015} found a strong correlation between the slope of a star cluster's stellar mass function and the fraction of mass lost by the cluster, $\floss$ (see also \citealt{Baumgardt2023}). This correlation remains consistent across various initial masses, orbits, and initial sizes of the cluster. They also derived a relation between the present-day mass function slope and the $\floss$ of a cluster (mass function–initial mass relation). Utilizing the mass function values of 29 Galactic GCs measured by \citet{Ebrahimi2020}, we applied the mass function–initial mass relation to determine their $\floss$. \figref{fig:BH190-DW05-23-tot} illustrates the mean Galactocentric distances (semi-major axes) plotted against $\frem$ for these MW GCs. The colour coding represents the varying degrees of $\floss$, which range from 0.6 to 0.9. The $\frem$ values for the GCs were also sourced from \citet{Ebrahimi2020}. We utilized the Galactic GCs database developed by Baumgardt and Sollima\footnote{https://people.smp.uq.edu.au/HolgerBaumgardt/globular/} to extract the orbital parameters of the GCs. \figref{fig:BH190-DW05-23-tot} shows that GCs with lower $\RG$ exhibit higher levels of $\floss$, consequently yielding greater $\frem$.

To determine whether the simulated models with $\sigma_\mathrm{BH} = 190\kms$ can reproduce the $\frem$ of MW GCs, we derived the best fitting function for $\frem$ of the modelled clusters in dependence on $\log _{10}(\dens)$ and $\log _{10}(\RG)$ at the moment when they have lost 60, 70, 80, and 90 per\,cent of their mass since formation (see  Equations \ref{eq:general} and \ref{eq:a-b}). We then set the initial density to $\log _{10}(\dens)=3.4$, which is a lower limit for the initial density of the MW GCs, and tracked the evolution of $\frem\left(\RG,\log _{10}(\dens)=3.4 \right)$ as a function of $\RG$. The colour-coded curves in \figref{fig:BH190-DW05-23-tot} represent $\frem\left(\RG,\log _{10}(\dens)=3.4 \right)$ for modelled clusters at $\floss=0.6$, 0.7, 0.8, and 0.9.

\figref{fig:BH190-DW05-23-tot} demonstrates that the curves of $\frem\left(\RG,\log _{10}(\dens)=3.4 \right)$ are not in alignment with the corresponding $\floss$ zones of MW GCs, which are visually indicated by color-coded cubes. For instance, the dark remnant retained in the models at $\floss=0.8$ corresponds to the remnant fraction of MW GCs with $\floss=0.6$. For MW GCs that have lost 80 per\,cent of their initial mass and are situated within the range $0<\RG<10 \kpc$, 55-70 per\,cent of their mass is attributed to the dark remnants. Whereas, for the simulated models with a similar mass loss fraction and Galactocentric distance, less than 40 per\,cent of the mass is assigned to WD. The figure also displays the $\frem$ value for the simulated cluster with a mass of $9\times10^4 \Msun$ from \citet{Haghi2020}. Notably, even when tripling the mass of the simulated clusters, the $\frem$ of the models still differs significantly from the $\frem$ of MW GCs. Additionally, the figure demonstrates that our estimate of $\frem$ for the models remains consistent even with a threefold increase in model mass. This consistency indicates that our results can be reliably extrapolated to masses approaching the initial mass of MW GCs, providing a useful framework for understanding remnant fraction trends across a broader mass range.

One might posit that faint stars, such as brown dwarfs, which fall below the telescope detection limits and were excluded from our modeled cluster mass range, could contribute significantly to the dark mass. However, brown dwarfs, which have masses $0.013-0.075\Msun$ \citep{Chabrier2000,Burrows2001}, constitute merely $\sim3.5$ per cent of a cluster's total mass based on the Canonical initial mass function \citep{Kroupa2001}. In conclusion, our analysis indicates that the assumption of a high natal kick for BHs, fails to accurately reproduce the present-day $\frem$ of MW GCs.

%_%_%_%_%_%_%_%_%_%_%_%_%_%_%_%_%_%_%_%_%_%_%_%_%_%_%_%_%_%_%_%_%_
\subsection{Absence of BH natal velocity kicks}\label{sec:NKick}

When a substantial population of BHs remains within a cluster, they fail to attain energy equilibrium with low-mass stars through energy equipartition, resulting in their runaway segregation towards the cluster's centre. This phenomenon, identified as Spitzer instability \citep{spitzer1987}, prompts the formation of a BHSub in the central part of the cluster. The centrally segregated BHSub exhibits significant dynamical activity, facilitating the formation of numerous BH-BH binaries (BBHs) via three-body interactions within the dense stellar environment \citep{spitzer1987,heggie2003}. Indeed, BHs interact in a single close encounter and share their kinetic energy. During such encounters, the more massive BHs form a binary system, while the comparatively less massive BH absorbs the surplus kinetic energy released in the three-body encounter and is propelled into a higher, less bound orbit.

Through subsequent encounters between BBHs and single BHs, the binary systems undergo gravitational tightening as single BHs are scattered to higher orbits. The scattered BHs transfer gained kinetic energy to the entire stellar population through two-body interactions, resulting in the expansion of the cluster \citep{Mackey2007,Banerjee2017,Rostami2024,Rostami2024-II}. Due to dynamical friction, the scattered BHs ultimately return to the cluster's core and join the BHSub. Each encounter further hardens the BBHs and amplifies their recoil velocity. If the BBHs reach a sufficient level of tightness, during the next encounter, a significant amount of kinetic energy can be transferred to either the BBHs or a single BH, leading to expulsion from the cluster. This mechanism is responsible for the self-depletion of the BHSub from the clusters. As a result, the BHSub operates like a power source, pumping kinetic energy from the cluster's central region to its surrounding stellar population. This energy injection process persists until the BHSub is fully depleted.

\begin{figure*}
    \centering
    \includegraphics[width=1\linewidth]{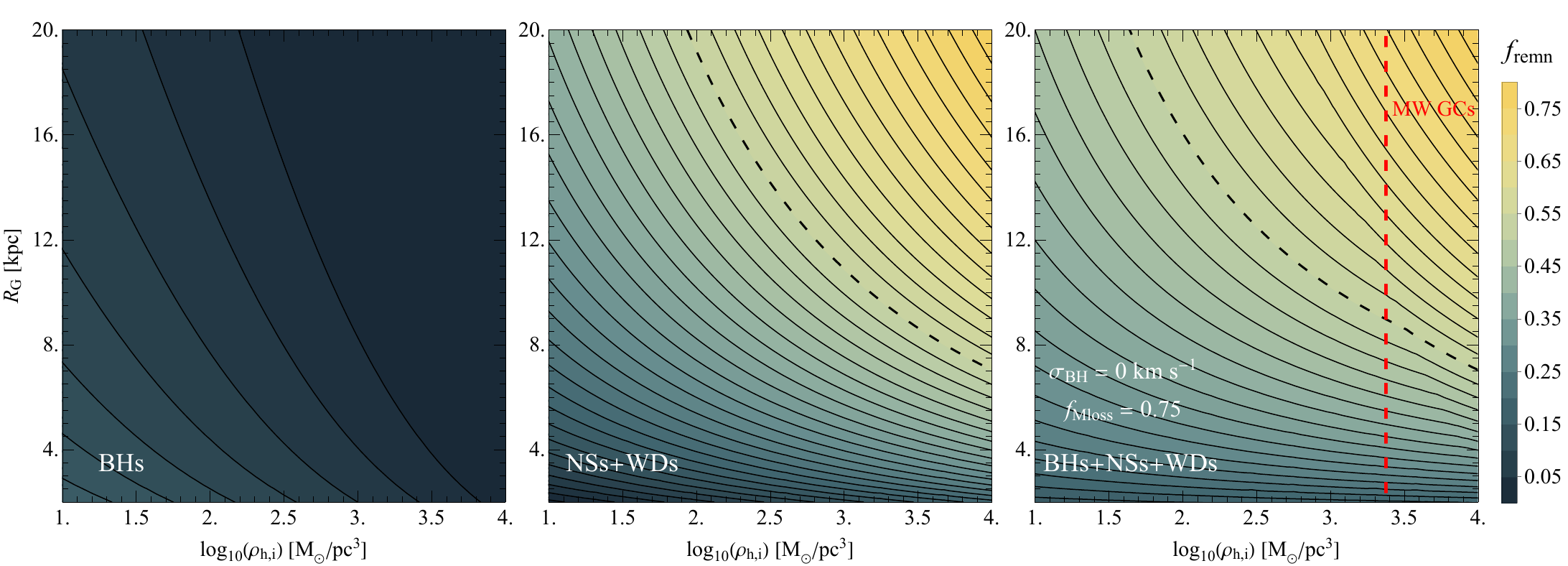}  
    \caption{Mass fractions of various stellar remnants in simulations featuring velocity dispersions of $\sigma_\mathrm{BH}=0\kms$ at an extracted snap shot of $\floss=0.75$. Left panel: mass fraction of BHs. Middle panel: combined NSs and WDs mass fraction. Right panel: their aggregated mass fraction.} 
    \label{fig:BH0-DW05-23-75per}
\end{figure*}

\begin{figure}
  \centering
  \includegraphics[width=0.9\linewidth]{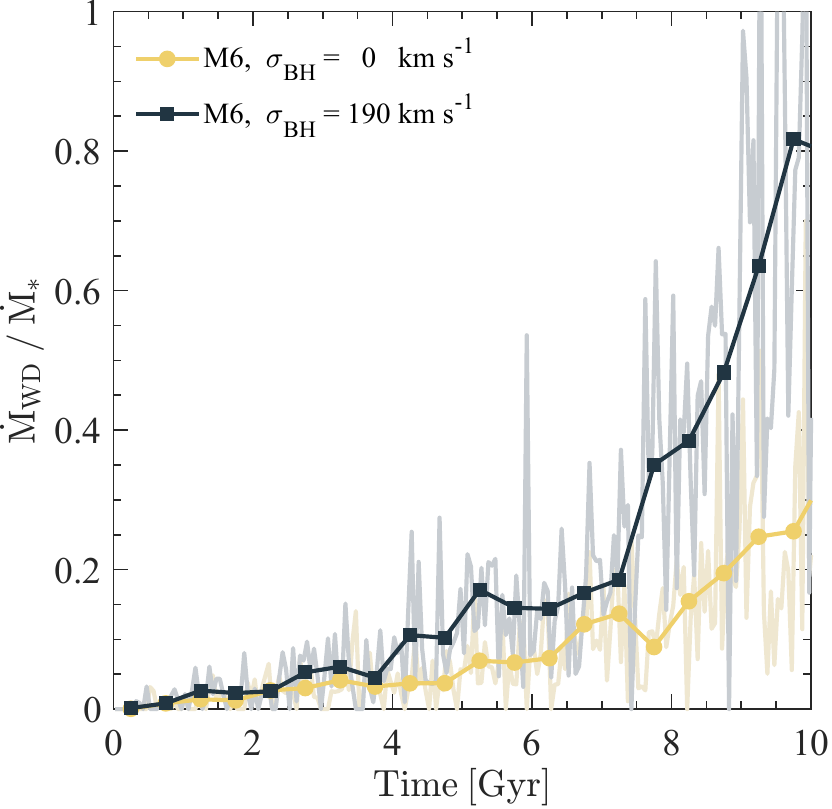}
  \caption{Similar to \figref{fig:dMWD_out_dMstar_time_rh06_rh2}, but comparing $\overset{.}{M}_\mathrm{WD} / \overset{.}{M}*$ of M6 models in simulations with different kick velocities. Yellow circles denote simulation with natal kicks of $\sigma_\mathrm{BH}=0\kms$, while black squares represent simulation with $\sigma_\mathrm{BH}=190\kms$.}
  \label{fig:dMWD-dMstar}
\end{figure}

In \citet{Rostami2024}, we elucidated how the initial density and $\RG$ of a star cluster influence the self-depletion rate of the BHSub. These parameters dictate whether the self-depletion time of BHSub surpasses the evaporation time of luminous stars or not. Indeed, they determine the survival of the BHSub throughout the cluster's lifetime. In this regard, the interplay between $\dens$ and $\RG$ is essential in explaining how the energy emitted by the BHSub influences the $\frem$ of the cluster. In the left panel of \figref{fig:BH0-DW05-23-75per}, the mass fraction of BHs ($\fBH$) is depicted in the extracted snapshot corresponding to $\floss=0.75$. The initial mass fraction of BHs predicted by stellar evolution is approximately $\fBH=0.04$. In clusters occupying regions with low $\RG$ and $\dens$, $\fBH$ increases over time, while dense clusters located at large $\RG$ experience a gradual reduction in $\fBH$ until $\fBH$ reaches zero.

During super-elastic encounters within the BHSub, a certain amount of kinetic energy is transferred to background stars. This released energy has the potential to propel luminous stars to escape velocity, thereby accelerating their evaporation rate. In dense star clusters, where the gravitational potential is deeper, the energy pumped from the BHSub might not be sufficient to propel luminous stars to escape velocity. On the other hand, before the evaporation of luminous stars, BHs are immersed in a deep central potential. During this period, the depletion rate of BHs is primarily contingent on the few-body encounters involving BBHs and solitary BHs, rather than on the tidal forces exerted by the host galaxy \citep{Breen2013,Longwang2020}. Actually, luminous stars in the cluster's halo serve as a shield against the galactic tidal force for the BHSub. Given that few-body encounters are more prevalent in denser clusters, the BHSub is depleted at a faster rate. Consequently, the self-depletion time of the BHSub is shorter than the evaporation time of luminous stars in dense clusters. The observed trend indicates that the $\fBH$ decreases as the initial density of the cluster increases.

Extending the $\RG$ of an orbiting cluster, which results in a weakened tidal field from the host galaxy, leads to a prolonged evaporation timescale. Conversely, the self-depletion timescale of the BHSub is primarily determined by the frequency of few-body encounters within the BHSub, rather than the tidal field, provided that the halo of luminous stars hasn't evaporated. Therefore, for a tidally underfilling cluster, the evaporation time of luminous stars increases with increasing $\RG$, whereas the self-depletion time of BHSub remains relatively constant. This leads to a reduction in $\fBH$ as Galactic radii increase. The left panel of \figref{fig:BH0-DW05-23-75per} reveals that for star clusters with low $\RG$ and $\dens$, $\fBH$ attains levels as high as 16  per\,cent. However, at $\log _{10}(\dens) > 4.1$, $\fBH$ tends towards zero. \figref{fig:BH0-DW05-23-75per} highlights that within the acceptable range of initial density for the MW GCs, the $\fBH$ is lower than 4 per\,cent. This fraction will diminish over the cluster evolution, gradually approaching zero. In \citet{Rostami2024}, we estimated the present-day $\fBH$ for MW GCs assuming a canonical initial mass function and full initial retention of BHs. Our investigation indicated that the majority of MW GCs are nearly depleted of BHs, with only 0-1 per\,cent of their total mass comprising BHs. However, around 13 per\,cent of MW GCs could still contain 1-4 per\,cent of their mass in BHs.

The middle panel of \figref{fig:BH0-DW05-23-75per} displays the cumulative mass fraction of NSs and WDs ($\fWD$). Remarkably, variations in $\fWD$ with respect to $\RG$ and $\dens$ exhibit an opposing trend to that of $\fBH$. The injection of energy from the BHSub into the background stars, driven by energy equipartition, tends toward a homogeneous distribution of kinetic energy among the stellar ensemble. Consequently, low-mass stars will experience greater velocity enhancements compared to higher-mass objects such as NSs or WDs. In clusters with low $\dens$ or $\RG$ parameters, signifying a low cluster escape velocity, the heightened velocity of NSs and WDs is sufficient to eject a notable portion of them from the cluster. The energy generated from BHSub within these clusters boosts both the evaporation rate of luminous stars and the escape rate of WDs/NSs, with no significant discrepancy in their expulsion rates. Thus, the $\fWD$ remains minimal in clusters characterized by low $\dens$ and $\RG$ (middle panel of \figref{fig:BH0-DW05-23-75per}).

Increasing the orbital radii and initial density of a cluster results in a corresponding increase in the cluster's escape velocity, rendering the energy generated from BHSub insufficient to accelerate WDs and NSs beyond the escape velocity. However, the velocity augmentation of low-mass luminous stars induced by the absorption of energy from BHSub enables them to accelerate beyond the escape velocity threshold, thereby increasing their evaporation rate. On the other hand, as BHSub undergoes self-depletion, the heaviest components of the cluster, i.e., NSs and WDs, segregate toward the cluster's centre and are protected from tidal stripping. Consequently, as the orbital radius and initial density of the cluster increase, the evaporation rate of luminous stars significantly surpasses the escape rate of WDs/NSs. This disparity leads to an increase in the $\fWD$ within the cluster. As depicted in the middle panel of \figref{fig:BH0-DW05-23-75per}, at $\RG = 20\kpc$ and $\log _{10}(\dens)=4$, the $\fWD$ can be as high as 76 per\,cent, indicating the predominance of WDs/NSs in the cluster.

In the right panel of \figref{fig:BH0-DW05-23-75per}, the mass fraction of all dark stellar remnants is depicted. Within the acceptable range for the initial density of the MW GCs, at $\RG>8\kpc$, $\frem$ exceeds 0.5, entirely attributed to the $\fWD$. Within this range, BHSub undergoes self-depletion, contributing nothing to $\frem$. Nevertheless, its initial retention plays a crucial role in substantially elevating $\frem$. Indeed, BHSub can be likened to a defunct factory. The energy emitted from this 'factory' accelerates the evaporation rate of luminous stars compared to the escape rate of WDs/NSs, thereby notably increasing $\frem$.

Note that the presence of NSs has a negligible impact on $\frem$, making them nonessential for the examination of star cluster dynamical evolution. We reconducted some simulations with $\sigma_\mathrm{BH}=0\kms$, assuming $\sigma_\mathrm{NS} = 190\kms$. These simulations revealed no significant change in $\frem$. \citet{Sollima2016}, analyzing simulations by \citet{Contenta2015}, concluded that despite the low mass fraction of NSs in clusters, they significantly contribute to the retention of WDs. It is noteworthy that in their simulations, the initial stellar mass range was constrained from 0.1 to 15 $\Msun$, excluding the presence of BHs in their modelled clusters. As a result, NSs were identified as the most massive remnants in their simulations. In more realistic simulations, the presence of BHs, as the heaviest stellar remnants, supersedes the influence of NSs in the dynamical evolution of star clusters, taking precedence in the narrative.

A comparison between Figures \ref{fig:BH190-DW05-23-75per} and \ref{fig:BH0-DW05-23-75per} indicates that in low initial density regimes, the models simulated with $\sigma_\mathrm{BH} = 190\kms$ exhibit higher values of $\frem$. The energy released by BHSub within this density regime is sufficient to kick out all components of the cluster, regardless of their mass, resulting in negligible discrepancies in their expulsion rates. Therefore, the inclusion of BHs results in a heightened rate of WD ejection relative to models lacking BHs.

However, this trend reverses in high-density ranges. In this regime, the escape rate of WDs in models with $\sigma_\mathrm{BH} = 190\kms$ escalates due to frequent few-body encounters within the concentrated WD population, initiating a gradual self-depletion process among the WD population. Conversely, the energy emitted by the BHSub within the modelled cluster, simulated with $\sigma_\mathrm{BH} = 0\kms$, induces a significant disparity in the ejection rates of WDs and luminous stars, thereby amplifying the evaporation rate of luminous stars. On the other hand, as BHs deplete from these modelled clusters, the segregation of WDs accelerates, thus promoting their survival against tidal stripping. In this case, the density of the segregated WD population proves insufficient to foster significant WD-WD binary formation. Indeed, the concentrated WD population is not dynamically active, thus avoiding self-depletion mechanisms.

The time-dependent evolution of $\overset{.}{M}_\mathrm{WD} / \overset{.}{M}_{*}$ is depicted in \figref{fig:dMWD-dMstar} for model situated at $\RG=12\kpc$ with $\rh=1\pc$ (model M6), across two different natal kick simulation series: $\sigma_\mathrm{BH} = 0\kms$ and $\sigma_\mathrm{BH} = 190\kms$. \figref{fig:dMWD-dMstar} indicates that $\overset{.}{M}_\mathrm{WD} / \overset{.}{M}_{*}$ is notably higher for models emptied of BHs through the application of a natal kick. This translates to a reduced remnant fraction of WDs, approximately 18 per\,cent lower compared to model simulated with $\sigma_\mathrm{BH}=0\kms$. Note that at $\mathrm{T}=10 \Gyr$, neither simulation retains any BHs within the clusters. Nevertheless, \figref{fig:dMWD-dMstar} demonstrates that the initial retention of BHs and their subsequent depletion through few-body encounters prove more successful in retaining WDs within the cluster compared to simulations where BHs were promptly ejected due to a natal kick.

\begin{figure}
  \centering
  \includegraphics[width=\linewidth]{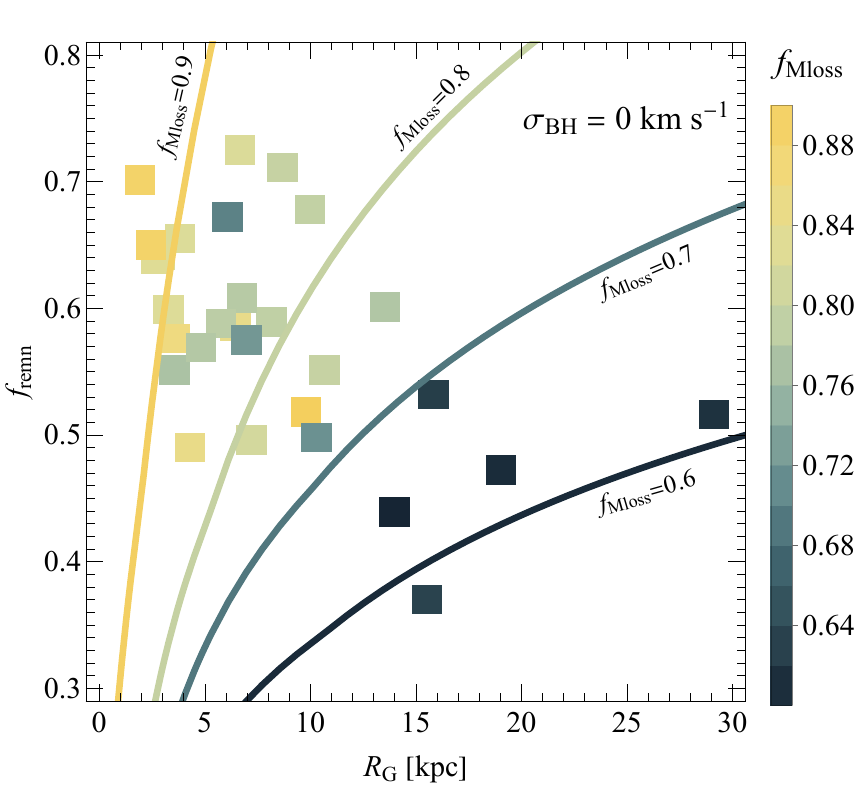}
  \caption{Same as \figref{fig:BH190-DW05-23-tot}, but for the simulation series of $\sigma_\mathrm{BH}=0\kms$.}
  \label{fig:BH0-DW05-tot}
\end{figure}

The color coded curves in \figref{fig:BH0-DW05-tot} show that, unlike the set of simulations with high BH natal kicks, $\frem$ of models simulated with low BH natal kicks accurately aligns with the dark mass fraction of MW GCs obtained from observations.

%%%%%%%%%%%%%%%%%%%%%%%%%%%%%%%%%%%%%%%%%%%%%%%%%%%%%%%%%%%%%%%%%%%%%%%%%%%%%
\section{DISCUSSION AND Conclusions}\label{sec:conclusion}

Comparing mass estimation inferred from the kinematics of stars within MW GCs with those derived from the conversion of light into mass reveals a more than two-fold discrepancy. Since this substantial invisible mass is segregated in the cluster core, the most likely hypothesis associates it with stellar remnants. The fraction of stellar remnants ($\frem$) within clusters is determined by the interplay of stellar evolution and dynamical processes. The majority of mass in remnants consists of WDs. Allocating over half of a GC's current mass to WDs could impose significant constraints on the dynamical evolution scenarios governing these stellar systems. As the heaviest element in GCs, populations of BHs play a crucial role in the dynamical evolution of GCs and exert a substantial effect on the escape rate of lower mass elements, such as WDs.

In this paper, we examined how BHs' natal kicks influence the fraction of dark stellar remnants ($\frem$) within star clusters. Our primary aim was to determine which of the evolutionary scenarios for the retention fraction of BHs could accurately reproduce the notable $\frem$ observed in MW GCs: 1) BHs receive low natal kicks, resulting in the formation of a subsystem concentrated within the central part of the cluster, driven by Spitzer instability; 2) A large fraction of BHs are ejected from the cluster immediately after formation due to a significant natal kick. Hence, we compared the $\frem$ of 29 MW GCs obtained by \cite{Ebrahimi2020} with a comprehensive grid of direct \Nbody simulations of clusters covering diverse initial half-mass radii ($\rh$) and Galactocentric distances ($\RG$), while adjusting the natal kicks received by BHs. Our simulations involved a range of BH kick velocities drawn from a Maxwellian distribution with a one-dimensional dispersion of $\sigma_{\mathrm{BH}}=0$ and $190\kms$ \citep{Hansen1997}, aiming to cover a wide range of possibilities for the mass fraction of BH remnants. In total, for our modelled clusters, we conducted two series of simulations based on varying natal kicks applied to BHs.

The determination of which natal kick scenario results in modelled clusters with the highest dark remnant fraction is contingent upon the initial density of the clusters, leading to distinct outcomes between low and high-density environments. In the high initial density regime, widely acknowledged as the acceptable range for the MW GCs, only simulations featuring minimal natal kicks to BHs succeeded in accurately reproducing the remnant fraction observed in MW GCs. According to the Spitzer instability, the presence of a substantial BH population within a cluster hinders the achievement of energy equilibrium with low-mass stars through energy equipartition. This disruption prompts the runaway segregation of BHs toward the cluster's centre, ultimately giving rise to the formation of a BHSub in the central part of the cluster. Within the cluster, the BHSub is a source of dynamic activity, continually generating kinetic energy through few-body encounters among BHs. In essence, it functions akin to an energetic power plant, transferring kinetic energy from the cluster's central region to its entire stellar population. This process continues until the BHSub is fully depleted from the cluster.

The injection of energy from the BHSub into the background stars, driven by energy equipartition, tends toward a homogeneous distribution of kinetic energy among the stellar ensemble. Consequently, luminous stars, possessing lower masses than WDs, undergo more pronounced velocity enhancements. This heightened velocity accelerates luminous stars beyond the escape velocity, subsequently elevating their evaporation rate. On the other hand, as BHSub undergoes self-depletion, the WDs segregate towards the cluster's centre, thus promoting their survival against tidal stripping. As a result, the energy emitted by the BHSub induces a notable discrepancy in the ejection rates of WDs and luminous stars, thereby contributing to an increase in the mass fraction of WDs within the cluster. Note that BHSub undergoes self-depletion during cluster evolution, contributing nothing to $\frem$. Nevertheless, the initial retention of BHs and their subsequent depletion through few-body encounters plays a crucial role in reproducing the observed $\frem$ of MW GCs. Indeed, BHSub can be likened to a defunct factory, whose existence was essential in the cluster's initial evolutionary stages.

This is the second paper in a series that delves into the necessity of the initial retention of BHs within GCs to replicate observational evidence. In \citet{Rostami2024-II}, we revealed that a high initial retention fraction of BHs can explain the observed dichotomies between metal-poor and metal-rich GCs. Moreover, in \citet{Rostami2024}, we demonstrated that achieving the present-day BH mass fraction observed in MW GCs does not require BHs to receive a high natal kick. Despite the high initial retention fraction of BHs, a considerable portion of them undergo depletion through few-body encounters, shaping the present-day BH mass fraction. Over the last decade, observations have confirmed the existence of BHs in star clusters. These findings imply that the natal kick received by BHs is not typically as large as previously thought. Moving forward, we aim to examine further observational evidence to assess whether employing low natal kicks for BHs can lead to a more refined alignment between simulations and observations of stellar clusters.

%%%%%%%%%%%%%%%%%%%%%%%%%%%%%%%%%%%%%%%%%%%%%%%%%%%%%%%%%%%%%%%%%%%%%%%%%
\section*{Acknowledgements}

This work is based upon research funded by Iran National Science Foundation (INSF) under project No.4035689.
%%%%%%%%%%%%%%%%%%%%%%%%%%%%%%%%%%%%%%%%%%%%%%%%%%%%%%%%%%%%%%%%%%%%%%%%%%%%%
\section*{Data availability}
The data underlying this article are available in the article.
%%%%%%%%%%%%%%%%%%%%%%%%%%%%%%%%%%%%%%%%%%%%%%%%%%%%%%%%%%%%%%%%%%%%%%%%%%%%%
% Bibliography

\bibliographystyle{mnras}
\bibliography{references}

%%%%%%%%%%%%%%%%% APPENDICES %%%%%%%%%%%%%%%%%%%%%
\appendix
\section{Isolating Dynamical Effects: Simulations with and without a Heavy Stellar Population}\label{sec:Appendix}

In this appendix, we present a series of simulations designed to isolate the dynamical effects of a heavy stellar population on cluster evolution, with a particular focus on the evaporation rates of lighter stellar populations. To disentangle the contributions of dynamical processes from stellar evolution, we disabled stellar evolution in these simulations, allowing us to focus purely on the dynamical evolution of clusters. This approach enables us to quantify the extent to which the observed trends in the mass fraction of WDs in clusters are driven by dynamical mechanisms versus stellar evolution processes.

We focused on the modeled cluster M5, characterized by an initial half-mass radius $\rh=1\pc$  and Galactocentric distance $\RG=8\kpc$. To explore the effects of varying $\rh$, we examined an additional model: M11, featuring a tripled initial half-mass radius ($\rh=3\pc$). For our analysis, we divided the stellar population into three distinct mass bins: $\mathrm{M_1}$ for stars with $\mathrm{m} \geq 8 \Msun$ (primarily BH progenitors), $\mathrm{M_2}$ for stars with $3 \Msun \leq \mathrm{m} < 8 \Msun$ (predominantly NS and WD progenitors), and $\mathrm{M_3}$ for stars with $\mathrm{m} < 3 \Msun$. We conducted two sets of simulations: one incorporating all mass bins to represent a complete stellar population, and another where we artificially removed the $\mathrm{M_1}$ population at the start of the simulation to mimic the effect of strong natal kicks on the most massive stars.

\begin{figure}

    \includegraphics[scale=0.57]{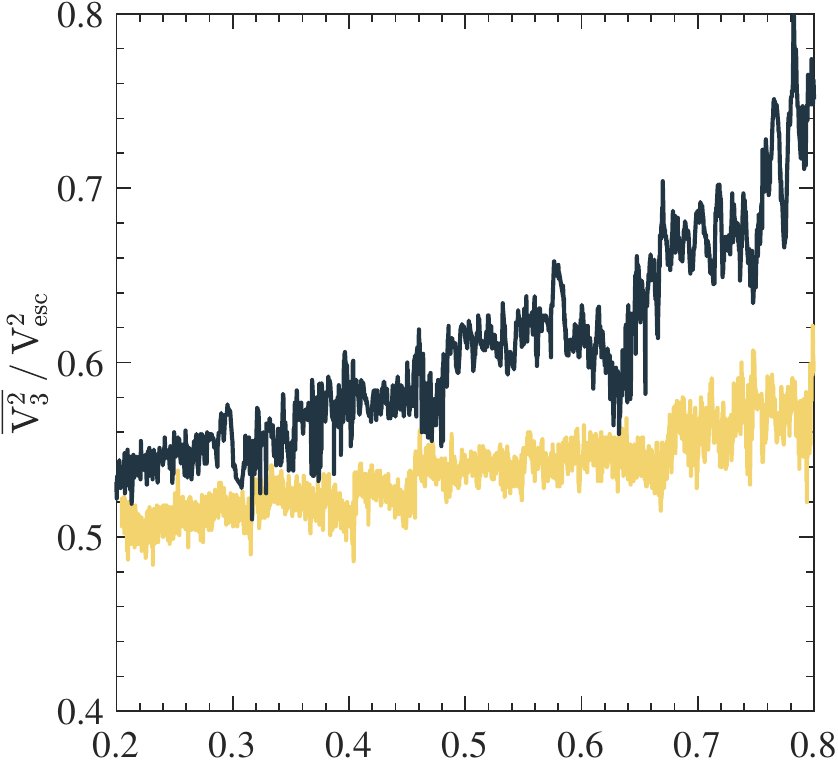}
    \includegraphics[scale=0.57]{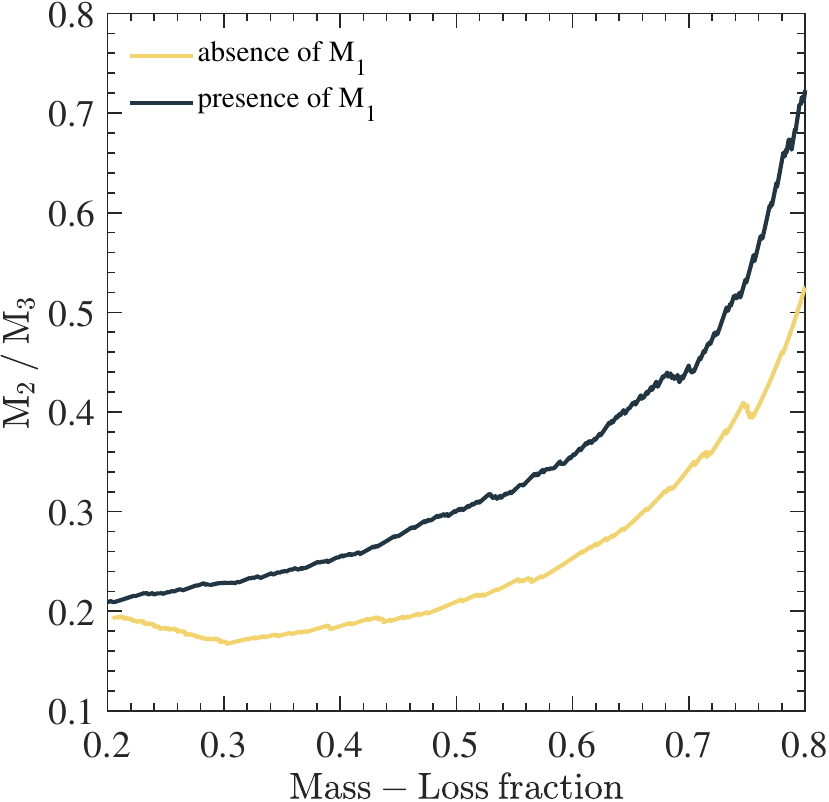}
 
    \caption{Evolution of dynamical parameters in simulated clusters with (black lines) and without (yellow lines) the heaviest stellar population ($\mathrm{M_1}$). Top panel: The ratio of mean square velocity of low-mass stars ($\mathrm{M_3}$) to the square of the cluster's escape velocity ($\overline{\mathrm{V_3^2}} / \mathrm{V}_\mathrm{esc}^2$) as a function of cluster mass loss fraction. Bottom panel: The mass ratio of intermediate-mass to low-mass objects ($\mathrm{M_2}/\mathrm{M_3}$) versus cluster mass loss fraction.}
    \label{fig:A1}
\end{figure}

The top panel of \figref{fig:A1} illustrates the evolution of $\overline{\mathrm{V_3^2}} / \mathrm{V}_\mathrm{esc}^2$, where $\overline{\mathrm{V}_3^2}$ is the mean square velocity of the $\mathrm{M_3}$ population and $\mathrm{V}_\mathrm{esc}$ is the cluster's escape velocity for both sets of simulations. While this ratio is initially 0.5 in a virialized cluster, it increases more rapidly in simulations including $\mathrm{M_1}$ stars, indicating a higher escape rate for the low-mass population. The bottom panel of \figref{fig:A1} illustrates the evolution of the $\mathrm{M_2}/\mathrm{M_3}$ mass ratio. Clusters retaining the $\mathrm{M_1}$ population maintain a consistently higher $\mathrm{M_2}/\mathrm{M_3}$ ratio over time compared to simulations where the $\mathrm{M_1}$ population is ejected. This trend suggests enhanced retention of intermediate-mass objects relative to their low-mass counterparts in the presence of the heaviest stellar population. This phenomenon can be attributed to energy equipartition: the energy generated by the $\mathrm{M_1}$ population in the cluster's center is preferentially transferred to $\mathrm{M_3}$ stars, accelerating their escape, while $\mathrm{M_2}$ stars experience less velocity enhancement. Consequently, the presence of $\mathrm{M_1}$ stars leads to a more pronounced disparity in escape rates between $\mathrm{M_2}$ and $\mathrm{M_3}$ populations, favoring the retention of intermediate-mass objects.

\begin{figure}
  \centering
  \includegraphics[width=\linewidth]{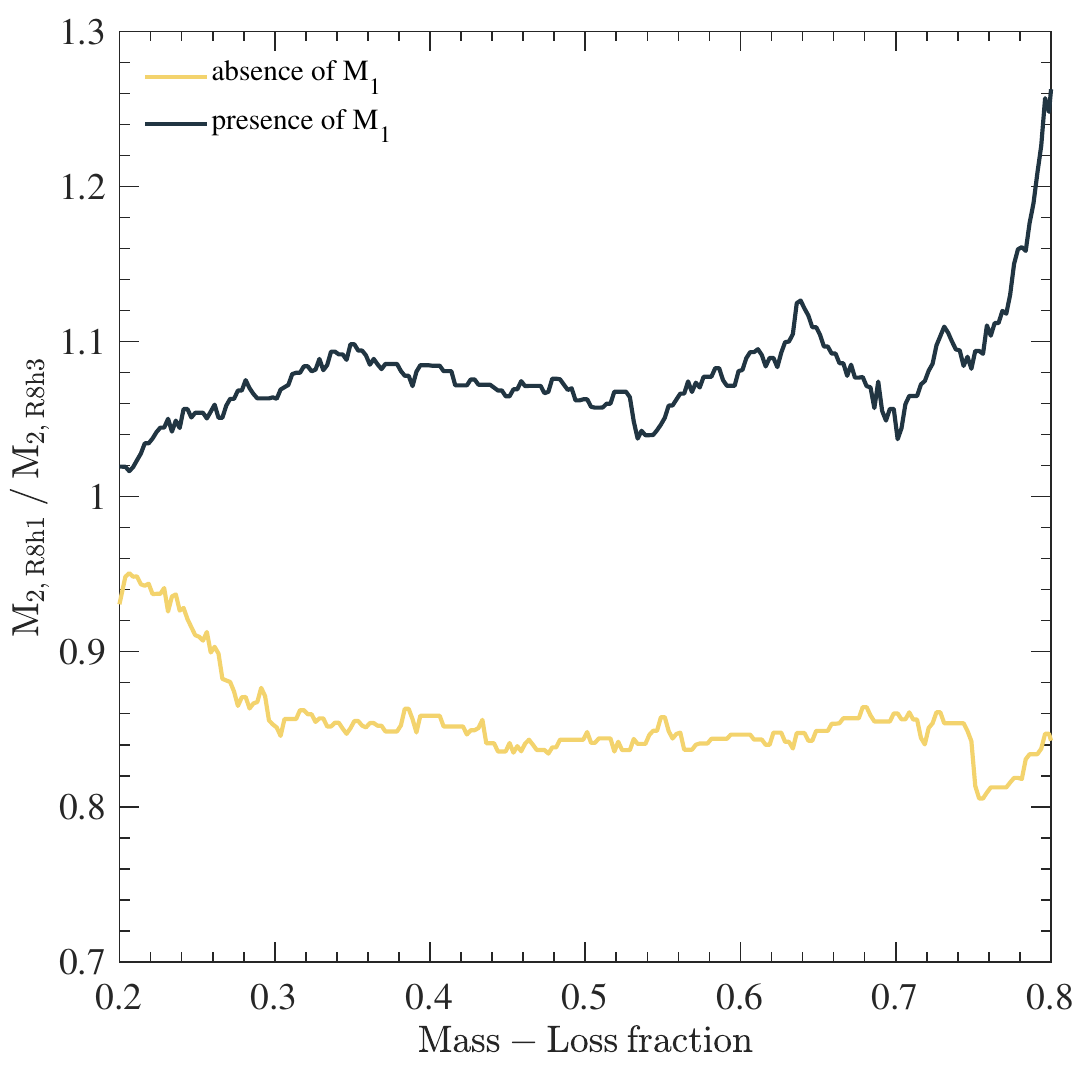}
  \caption{The ratio of retained $\mathrm{M_2}$ population in clusters with initial half-mass radius of 1 pc to those with 3 pc ($\mathrm{M}_\mathrm{2, R8h1} / \mathrm{M}_\mathrm{2, R8h3}$). Black lines represent simulations including all mass bins, while yellow lines show simulations where the $M_1$ population is ejected at the start.}
  \label{fig:A2}
\end{figure}

\figref{fig:A2} presents the evolution of the $\mathrm{M_2}$ population for simulations with varying initial half-mass radii (1 and 3 pc). In simulations where $\mathrm{M_1}$ stars are retained, clusters with higher initial density (characterized by smaller $\rh$) exhibit enhanced retention of intermediate-mass objects ($\mathrm{M}_\mathrm{2, R8h1} > \mathrm{M}_\mathrm{2, R8h3}$). The accelerated self-depletion of the $\mathrm{M_1}$ population in dense environments drives this phenomenon. As the $\mathrm{M_1}$ population diminishes, $\mathrm{M_2}$ objects undergo mass segregation towards the cluster center, where they become less susceptible to tidal stripping effects. Conversely, in simulations lacking $\mathrm{M_1}$ stars, the $\mathrm{M_2}$ population directly undergoes mass segregation to the cluster center, becoming dynamically active through frequent few-body encounters. Under these conditions, higher initial cluster density can lead to expedited self-depletion of $\mathrm{M_2}$ objects due to an increased frequency of dynamical interactions ($\mathrm{M}_\mathrm{2, R8h1} < \mathrm{M}_\mathrm{2, R8h3}$).

These simulations illuminate the pivotal role of massive stars in governing the retention of stellar remnants within clusters. The presence of a heavy stellar population substantially alters the differential evaporation rates among various mass groups, leading to enhanced retention of intermediate-mass objects even after the depletion of the most massive stars. Our findings demonstrate that dynamical processes exert a dominant influence on the stellar remnant fraction, significantly outweighing the effects of stellar evolution. The trends in WD mass fraction elucidated in the main text can be predominantly ascribed to intra-cluster dynamical mechanisms rather than stellar evolutionary processes.

\bsp
\label{lastpage}
\end{document}